# Ga-rich GaAs(001) surfaces observed during high-temperature annealing by scanning tunneling microscopy

Shiro Tsukamoto[a], Markus Pristovsek, Bradford G. Orr[b], Akihiro Ohtake, Gavin R. Bell[c], and

Nobuyuki Koguchi

National Institute for Materials Science

1-2-1 Sengen, Tsukuba, Ibaraki 305-0047, JAPAN

## ABSTRACT

Ga-rich GaAs (001) surfaces are successfully observed during high-temperature annealing by scanning tunneling microscopy (STM). With a substrate temperature of 550 $^{o}$C, reflection high-energy diffraction patterns and reflectance anisotropy spectra confirm a (4x2) Ga-stabilized surface. STM images clearly show alteration of the surface reconstructions while scanning. It is postulated that detaching and attaching of Ga adatoms may be the cause of these surface dynamics. For these reasons it is determined that $\zeta$(4x4), $\zeta$2(4x4) and $\zeta$(4x6) reconstructions co-exist on the surface. The $\zeta$2(4x4) reconstruction contains a Ga tetramer cluster and in more Ga-rich conditions, the $\zeta$2(4x6) surface has a Ga octamer cluster.

a) e-mail: TSUKAMOTO.Shiro@nims.go.jp

b) present address: The University of Michigan, Ann Arbor, MI48109, U.S.A.

c) present address: University of Warwick, Coventry, CV4 7AL, U.K.

# Ga-rich GaAs(001) surfaces observed during high-temperature annealing

# by scanning tunneling microscopy

Shiro Tsukamoto[a)], Markus Pristovsek, Bradford G. Orr[b)], Akihiro Ohtake, Gavin R. Bell[c)], and

Nobuyuki Koguchi

National Institute for Materials Science

1-2-1 Sengen, Tsukuba, Ibaraki 305-0047, JAPAN

The complexity and variety of surface reconstructions observed on semiconductors has been an intriguing problem in surface science. Recently, combined molecular beam epitaxy (MBE) and scanning tunneling microscopy (STM) systems have provided a very powerful technique for the real-space observation of semiconductor surfaces, especially GaAs(001), with extremely high resolution[1-5]. Avery *et al.* reported STM studies of MBE grown submonolayer islands in the pre-coalescence regime on the three low-index surfaces of GaAs[6,7]. In Ref.7, the dynamics of the As-rich surface during growth were inferred by comparing room-temperature STM 'snapshots' with kinetic Monte Carlo simulations[7]. Ga-rich surfaces[8-10] are especially difficult to study due to a transition in surface reconstruction between high and low temperatures[11]. Therefore, if samples are cooled and transferred to a cleaner environment for STM analysis, the surfaces are no longer representative of the one at high temperature. STM analysis at high-temperature needs to be performed. In this paper, the Ga-rich GaAs (001) surfaces were successfully observed during high temperature annealing by STM.

Si-doped GaAs(001) $1^{o}$ off <111>A (n = $2x10^{18}$ cm$^{-3}$) substrates were prepared by standard solvent cleaning and etching procedures, and then loaded into the MBE chamber. After the oxide was removed at 600 $^o$C under an As$_4$ flux of 2x10$^{-5}$ torr, an undoped GaAs buffer layer of 1.0 μm was grown at 580 $^o$C with the following conditions: As$_4$/Ga flux ratio ~30, growth rate 1.0 μm/hour and background pressure of 2x10$^{-7}$ torr. This resulted in a smooth GaAs surface with single bilayer steps. The samples were then annealed at 550$^o$C without As overpressure in order to produce a (4x2) phase surface as determined by reflection high-energy electron diffraction (RHEED)[12,13], as shown in Fig.1(a). The background pressure was 2x10$^{-11}$ torr. Under the same conditions reflectance anisotropy spectra (RAS or RDS) and STM images were obtained.

The RAS spectrum, shown in Fig.1(b), indicates a (4x2) Ga-stabilized surface. The STM images show steps and a few islands present on the surface. Figure 2 shows the changes in the surface reconstruction observed by STM. Line scans and correlation functions of the surface confirm the clear 4x periodicity in the [1-10] direction and show regions of the surface with a periodicity of x3 and x4 in the [110].

What are the possible surface reconstructions that are consistent with RHEED analysis, RAS measurement and the STM images? The STM data strongly points to a co-existence of reconstructions. One set of candidates is the ζ(4x4), ζ(4x6), and ζ2(4x4) reconstructions, as shown in Fig.3(a), (b), and (c) respectively. All models are based on ζ(4x2) by Lee *et al.*[8] and satisfy electron-counting heuristics. The models differ in the presence and location of Ga atoms. At elevated temperatures Ga adatoms can detach and diffuse to make Ga clusters[14]. These dynamics may be the cause for the changes shown in Fig.2. Mobile Ga would result in different surface reconstructions on different parts of the surface as seen in Fig.2.

In the ζ(4x4) reconstruction shown in Fig.3(a), 50% of the Ga dimers in ζ(4x2) are missing and four As dangling bonds make two As dimers. Therefore, the STM images show that the trench is unfilled as indicated by a white arrow in Fig.2(i). The $\zeta2(4x4)$ reconstruction, shown in Fig.3(c), does not make As dimers but depends on a Ga tetramer magic cluster that consists of 4 adatoms and is able to supply 4 electrons to the four As dangling bonds[15]. The STM image of this reconstruction would show that the trench is partially filled with the Ga clusters as indicated by white arrow in Fig.2(g). The $\zeta(4x6)$ reconstruction, shown in Fig.3(b), has two out of three Ga dimers missing as well as two As adatoms in each missing Ga dimer region. After restructuring, two Ga and two pairs of As form new dimers. This surface leads to STM images that show dark lines every three lattices units along [110] as indicated by white arrow in Fig.2(h). Each of these reconstructions does not form large domains and is distributed randomly on the (001) surface. This reasonably explains why we do not observe the 1/4- and 1/6- order reflections in RHEED patterns obtained along the [1-10] direction. However, since all reconstructions are derived from the $\zeta(4x2)$, the surface dynamics associated with Ga motion[14] will produce transient regions with this symmetry. Therefore, it is natural to observe the 1/2- order reflection along the [1-10] direction.

From STM images, we can determine the average distribution of the $\zeta(4x4)$, $\zeta(4x6)$, and $\zeta2(4x4)$ reconstructions, which are 61%, 19%, and 20%, respectively. For this mixed surface structure the occupancies of Ga dimers and cluster sites are determined to be 47% and 20%, respectively. The average coverage of surface Ga adatoms is 0.67 ML. These values are consistent with the results determined by surface x-ray diffraction using direct methods by Kumpf *et al*.[10] and by rocking-curve analysis of RHEED[16].

For the more Ga-rich case, $\zeta2(4x6)$ surface, which contains a Ga octamer magic cluster, is predicted to be formed as shown in Fig.3(d). This Ga octamer is also able to supply 4 electrons to the As dangling bonds to satisfy electron counting[15]. The $\zeta2(4x6)$ is not observed in our samples, but matches very well with the Ga dosed (4x6) surface observed by Xue *et al*. with STM [13,17]. In the $\zeta2(4x6)$ model, the occupancies of Ga dimers and cluster sites are estimated as 67% and 133%, respectively. Therefore, the coverage of surface Ga adatoms is estimated as 1 ML, which is also very consistent with the results by Xue *et al*.[17].

In summary, we have use high-temperature STM, RHEED and RAS to examine the GaAs (001) surface. We find the surface to contain several different coexisting reconstructions. These reconstructions are all derived fundamentally from the $\zeta(4x2)$ structure. The mixed surface is consistent with electron counting rules and also agrees with X-ray and RHEED rocking-curve data. It will be very instructive to examine the stability of these reconstructions, in particular those with closed-shell Ga clusters, by density functional total energy calculations[18].

The authors wish to thank Dr.T.Ohno, NIMS, for his instructive discussions and helpful comments. This study was partially performed through Special Coordination Funds of the Ministry of Education, Culture, Sports, Science and Technology of the Japanese Government.

**References**

1. D.K.Biegelsen, R.D.Bringans, J.E.Northrup, and L.E.Swartz, Phys.Rev. **B41**, 5701 (1990).

2. I.Tanaka, S.Ohkouchi, T.Kato, and F.Osaka, J.Vac.Sci.&Technol. **B9**, 2277 (1991).

3. B.G.Orr, C.W.Snyder, and M.Johnson, Rev.Sci. Instrum. **62**, 1400 (1991).

4. M.Tanimoto, J.Osaka, T.Takegami, S.Hirono, and K.Kanisawa, Ultramicroscopy **42/44**, 1275 (1992).

5. T.Hashizume, Q.Xue, J.Zhou, A.Ichimiya, and T.Sakurai, Phys.Rev.Lett. **73**, 2208 (1994).

6. A.R.Avery, H.T.Dobbs, D.M.Holmes, B.A.Joyce, and D.D.Vvedensky, Phys.Rev.Lett. **79**, 3938 (1997).

7. M.Itoh, G.R.Bell, A.R.Avery, T.S.Jones, B.A.Joyce, and D.D.Vvedensky, Phys.Rev.Lett. **81**, 633 (1998).

8. S.-H.Lee, W.Moritz, and M.Scheffler, Phys.Rev.Lett. **85**, 3890 (2000).

9. W.G.Schmidt, S.Mirbt, and F.Bechstedt, Phys.Rev.B **62**, 8087 (2000).

10. C.Kumpf, D.Smilgies, E.Landemark, M.Nielsen, R.Feidenhans'l, O.Bunk, J.H.Zeysing, Y.Su, R.L.Johnson, L.Cao, J.Zegenhagen, B.O.Fimland, L.D.Marks, and D.Ellis, Phys.Rev.B **64**, 75307 (2001).

11. I.Kamiya, D.E.Aspnes, L.T.Florez, and J.P.Harbison, Phys.Rev.B **46**, 15894 (1992).

12. S.Tsukamoto and N.Koguchi, Jpn.J.Appl.Phys., **33**, L1185 (1994); J.Cryst.Growth **150**, 33 (1995).

13. Q.Xue, T.Hashizume, J.M.Zhou, T.Sakata, T.Ohno, and T.Sakurai, Phys.Rev.Lett. **74**, 3117 (1995).

14. S.Tsukamoto and N.Koguchi, Materials Research Society Symposium Proceedings **648**, P11.20 (2001).

15. S.Tsukamoto and N.Koguchi, J.Cryst.Growth **209**, 258 (2000); J.Cryst.Growth **201/202**, 118 (1999).

16. A.Ohtake, S.Tsukamoto, N.Koguchi, and M.Ozeki, unpublished.

17. Q.Xue, T.Hashizume, and T.Sakurai, Progress in Sur. Sci. **56**, 64 (1997).

18. H.Hakkinen and M.Manninen, Phys.Rev.Lett. **76**, 1599 (1996); J.Chem.Phys. **105**, 10565 (1996).

**Figure Captions**

Figure 1. (a) RHEED pattern and (b) RAS spectra of Ga-rich GaAs (001) surface with the substrate temperature of 550 $^o$C. For comparison, RAS spectra of the As-rich c(4x4), (2x4), and (nx6), or mixed (2x6) and (3x6), surfaces are also shown.

Figure 2. (a) to (i) a series of STM images of Ga-rich GaAs (001) surface during annealing with the background pressure of $2 \times 10^{-11}$ torr. Images were obtained in constant current mode using a sample bias of -3.5V (filled states) and tunneling currents of 0.2nA. An island with lateral dimension of 2nm, shown as white in the images, was used as a marker that does not change its position during scanning. Therefore, the images are drifting along [-1-10] direction.

Figure 3. Top and side views of (a) ζ(4x4), (b) ζ(4x6), (c) ζ2(4x4) and (d) ζ2(4x6) surface reconstruction models.

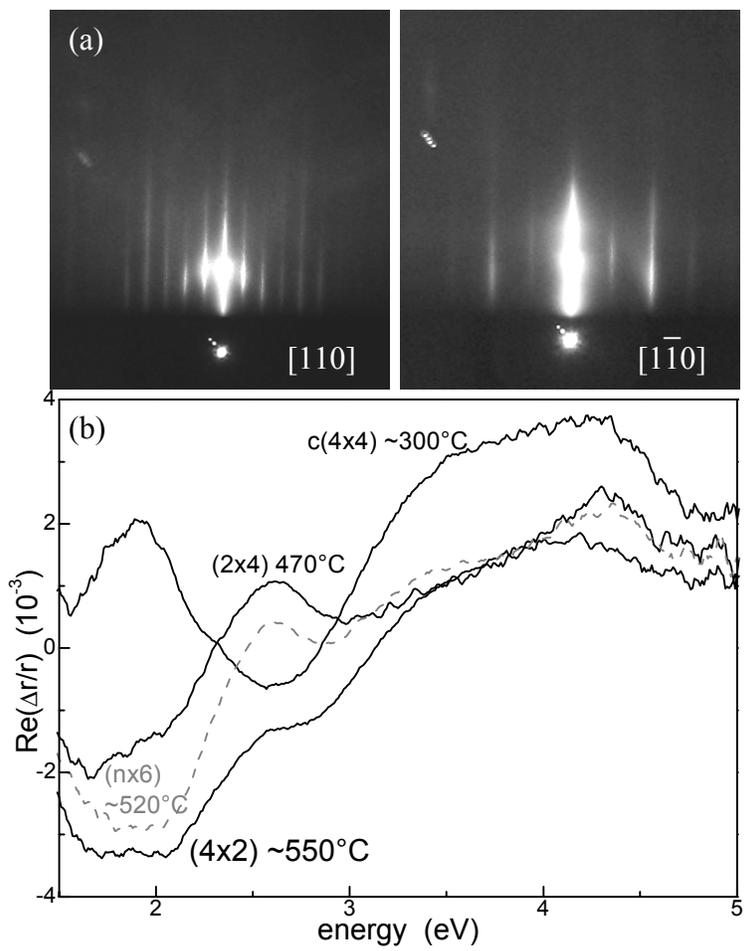

**Figure 2  Tsukamoto at al.**

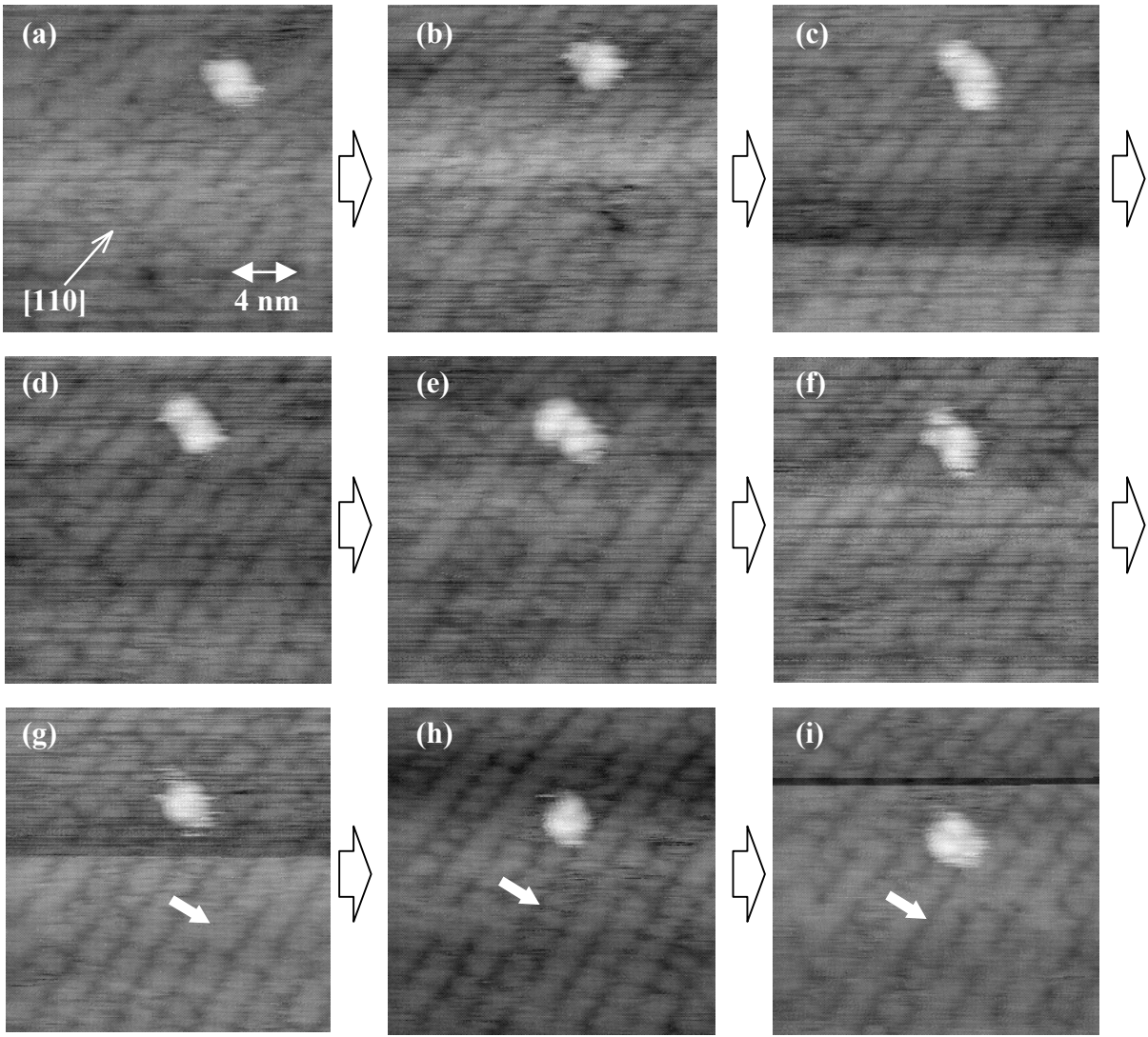

**Figure 3  Tsukamoto at al.**

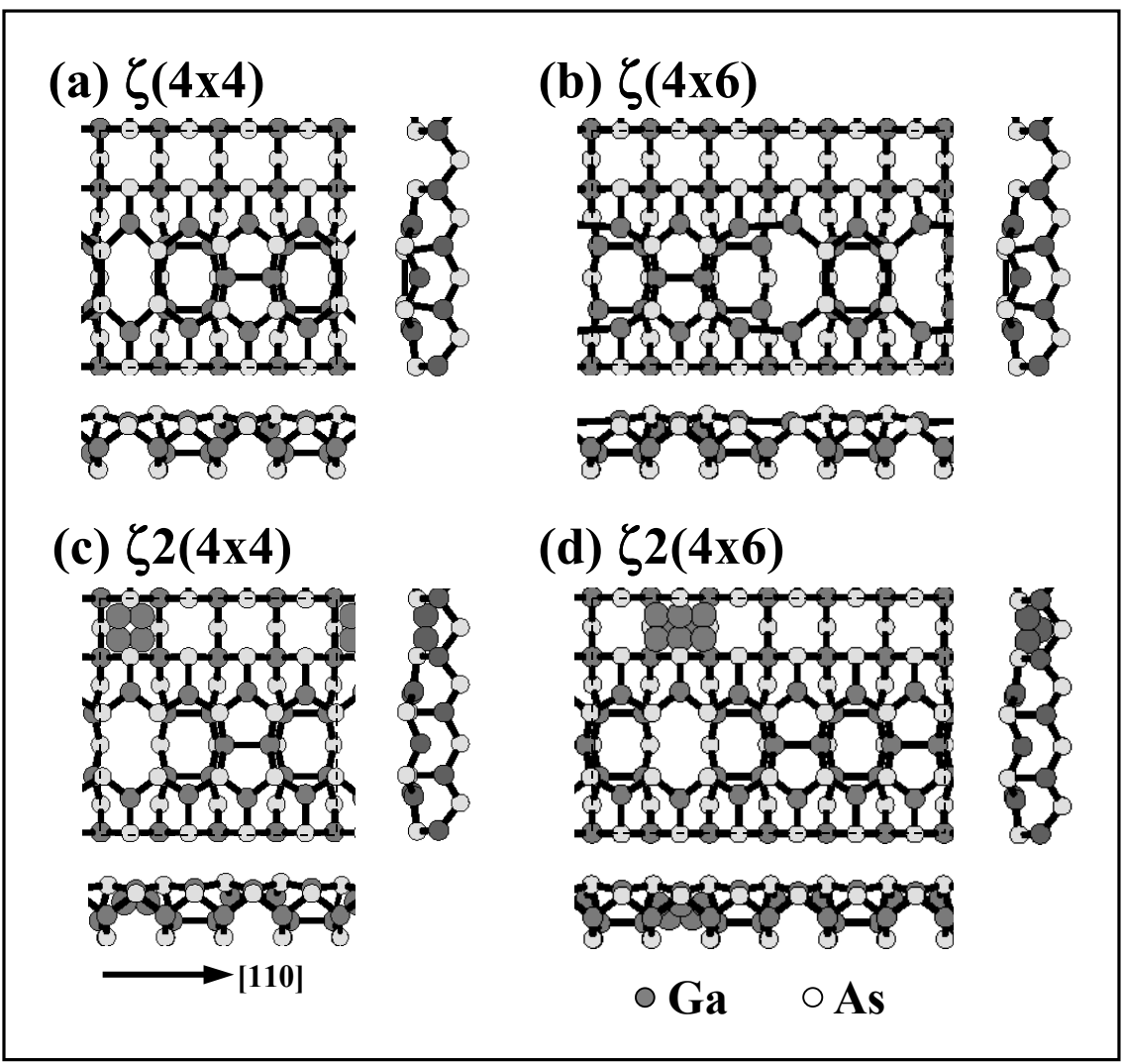

(a) ζ(4x4)     (b) ζ(4x6)

(c) ζ2(4x4)    (d) ζ2(4x6)

[110]

● Ga    ○ As